\documentclass[prl,aps,twocolumn]{revtex4}
\usepackage[ansinew]{inputenc}
\usepackage{amssymb}
\usepackage{graphicx}
\usepackage{amsmath}

\begin{document}
\title{Energy-time entanglement preservation in plasmon-assisted light transmission}
\author{Sylvain Fasel}
\email{sylvain.fasel@physics.unige.ch}
\author{Nicolas Gisin, Hugo Zbinden}
\affiliation{Group of Applied Physics, University of Geneva\\
CH-1211 Geneva 4, Switzerland}
\author{Daniel Erni, Esteban Moreno, Frank Robin}
\affiliation{Communication Photonics Group, ETHZ\\
 8092 Zürich, Switzerland}

\begin{abstract}
We report on experimental evidences of the preservation of the energy-time entanglement for extraordinary plasmonic light transmission through sub-wavelength metallic hole arrays, and for long range surface plasmon polaritons. Plasmons are shown to coherently exist at two different times separated by much more than the plasmons lifetime. This kind of entanglement involving light and matter is expected to be useful for future processing and storing of quantum information.
\end{abstract}
\maketitle

Entanglement is one of the most fundamental aspect of the quantum theory. While it is somewhat counter-intuitive and has no classical counterparts, it can be used to achieve information and communication tasks with much higher efficiencies than otherwise possible classically, and is thus the heart of the quantum information. A particular form of entanglement, the energy-time entanglement involving photons at telecom wavelength is particularly efficient in carrying quantum information over large distances as it has been shown to be specially robust against environmental perturbation \cite{robust}. This already leads to applications like quantum cryptography \cite{cryptoreview} or quantum teleportation \cite{teleportbase}. However, entanglement involving distant solid matter is a mandatory but difficult step for useful quantum information processing and storing, and researches are held in that direction \cite{monroe,polzik2,barrett,polzik3}. Interaction between energy-time entangled photons at telecom wavelength and quantum states of matter is thus of great interest, both from a fundamental point of view and for its potential future applications.

We investigate the coupling of energy-time entangled photon pairs with surface plasmons (SPs), which are collective excitation wave involving typically $10^{10}$ free electrons propagating at the surface of conducting matter (usually metal) \cite{plasmonsreview}. To this end, we placed gold films perforated with periodic subwavelength hole array in the path of energy-time entangled photons. Since the holes are subwavelength, this kind of nanostructures impedes the direct photon transmission, but allows the resonant excitation of a SP at the metal interfaces, which reradiates a photon at the back side of the metallic film. The transmittance of this plasmon-mediated phenomenon is orders of magnitude larger than predicted by the standard theory of diffraction by small metallic apertures \cite{bethe}. This phenomenon, called extraordinary optical transmission \cite{extra,moreextra,moreextra2,theoextra} is known to preserve polarization entanglement under certain geometric conditions \cite{polentang,theoentang}.

In this letter we present evidence that energy-time entanglement
also survives this photon--plasmon--photon conversion with photons
at telecom wavelength. To this end we measure the strength of
non-local quantum correlations involving photons from entangled
pairs, using a Franson-type interferometric experiment
\cite{franson}. We measured it in the case that both photons go
directly to the interferometers, and in the case that one member
of the entangled pair undergoes a plasmon conversion before
reaching its interferometer. The strength of the quantum
correlation is quantified by means of the visibility of
interferences fringes, recorded in both cases. We show that no
visibility reduction is observed and thus that entanglement in
preserved.

\begin{figure}[h]
\includegraphics[width=\columnwidth]{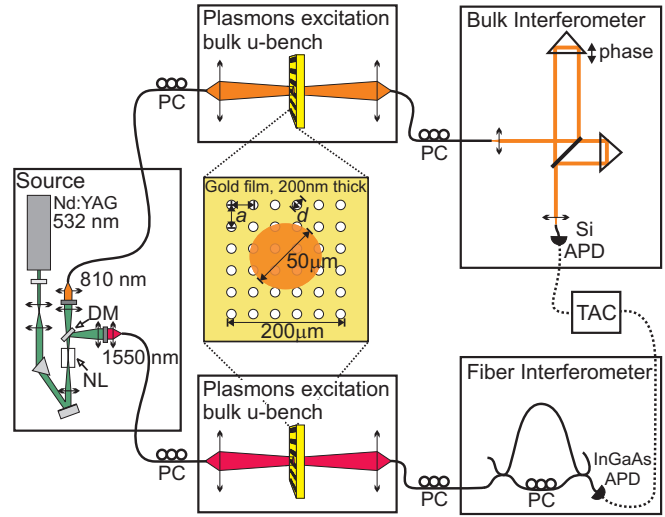}
\caption{Schematic of the experimental setup. NL: non-linear
crystal; DM: dichroic mirror; PC: polarization controller; $a$:
lattice periodicity, $d$: hole diameters (see text).}
\label{setup}
\end{figure}

The experimental setup is made of a photon pairs source, two
U-benches allowing the two different perforated gold film samples
to be inserted and removed in the paths of the photons, and two
interferometers (see figure \ref{setup}).
The photon pairs are created inside a KNbO$_3$ non-linear crystal
pumped with a 532\,nm NdYaG continuous laser (coherence length
$>1$ km) in a type I spontaneous parametric down conversion (SPDC)
configuration. One member of the pair has its central wavelength
around 810\,nm with a spectral width of about 2\,nm while the
other is centered around 1550\,nm with a spectral width of 7\,nm.
These values correspond to coherence times of 1.1\,ps and
coherence lengths of 0.34\,mm. Adjusting the phase matching by
tilting the crystal, it is possible to tune slightly these central
wavelength. The SPDC generated photons are collected into two
single mode fibers by mean of a dichroic mirror and two coupling
lenses and sent to the U-benches. As this source produces photons
at two well separated wavelengths, the properties of two different
gold subwavelength nanostructures are investigated successively.
One is designed to match the 810\,nm photons the other the
1550\,nm photons. The photons that undergo the plasmonic
conversion are either the 810\,nm one or the 1550\,nm ones.

The U-benches are made of two lenses. The first focuses the light
at the output of the first single mode fiber into a beam with a
50\,$\mu$m beam diameter situated a few centimeters away from the
lens. The second lens couples back this beam inside another single
mode fiber that is finally connected to an interferometer.
Perforated gold film samples can be hold at the beam waist
position  perpendicularly to the beam optical axis, by an
orientable mount. Without samples, the U-benches feature about
3\,dB insertion losses, without measurable spectral or
polarization dependency. When a sample is inserted into the photon
beam, no corrections are made to the lense's alignement. The light
that is collected back into the fiber is therefore the light which
is re-emitted in the same spatial mode than in the case where no
sample is inside the beam. Control of the incident polarization
state is made using the fiber polarization controller in front of
the U-benches. However, as expected in this particular square hole
array configuration for which one can assume that photons arrive
perpendicularly to the sample (less than 1 degree of angular
spreading), the polarization dependency of the transmittance
$T_\lambda$ at the operating wavelength $\lambda$ was observed to
be less than 2\,dB. The other polarization controllers are used to
control the polarization at the input of both interferometers and
inside the fiber interferometer in order to maximize the optical
visibility.

The output of the U-benches are connected to unbalanced
Mach-Zender interferometers, with 1\,m optical path length
difference between the two arms (this correspond to a time
difference of 3.3\,ns). The one that is designed for the 810\,nm
photons is made out of bulk optics, while the other is made out of
fibers. Both interferometers are passively stabilized by means of
temperature regulation. 810\,nm photons are detected at one output
of the bulk interferometer using an actively quenched silicon APD
photon counter (EG\&G). The 1550\,nm photons are detected at one
output of the fiber interferometer using one InGaAs APD photon
counter (idQuantique), gated with the signal of the silicon
counter. The width of the gate applied on the the InGaAs APS is
about 2.5\,ns. The dark counts of this detector is about
$3.5\times10^{-5}$ counts per gate. The dark counts of the Si APD
is negligible. More details about this source and the
interferometers can be found in \cite{cryptoribordy,cryptofasel}.

Photons independently travel through the long arm and short arms
of their respective interferometers, and the time differences
between the detections of both photons are measured using a time
to analog converter (TAC). Photons have a 50\% probability of both
choosing the long arms or both the short arms of their respective
interferometers. In both these cases, the time difference between
the two detections is about zero. Since the coherence of the pump
laser is much larger than the interferometers's imbalance, it is
impossible to know which paths they have chosen. These events are
thus undistinguishable. This leads to quantum interferences that
correlate the chosen output port for both photons. Theses quantum
correlations are functions of the sum of the phases applied inside
the interferometers  and reflect the energy-time entanglement of
the photon pairs. By using a time window discriminator it is
possible to isolate these interfering events from the
non-interfering ones (i.e. photons choosing different arms) as the
latter happen at different times. In this experiment the sum of
the phases is adjusted by tuning the length of the long arm of the
bulk interferometer using a piezo-electric actuator. The
visibility of the  interference fringes is a direct indication of
the strength of the quantum correlation due to the energy-time
entanglement, and thus can be used to verify that the energy-time
entanglement is not destroyed by the photon--plasmon--photon
conversion.

The two different perforated film samples were fabricated in
200nm-thick evaporated gold films on 0.9\,mm glass substrates. The
gold was coated with a 200nm-thick PMGI (polydimethyl glutarimide)
electron-beam sensitive resist layer which was patterned using
electron-beam lithography. Following exposure of the hole arrays,
the resist was developed in tetraethylammonium hydroxide for 1min.
at 22°C and rinsed with de-ionized water. The structures were then
transferred into the Au layer using Ar-sputtering for 14 min. with
200W power. Finally, the PMGI residuals were removed using oxygen
ashing. The arrays size is about $200\times200\,\mu$m$^2$.  The
holes are circular and disposed following a regular square
pattern. As the light beam area is only about 5\% of the array
area, the array can be considered as infinite, the boundary
effects are thus negligible.

\begin{figure}[h]
\includegraphics[width=\columnwidth]{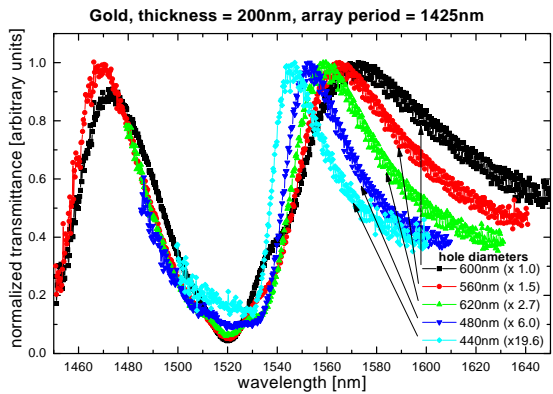}
\caption{Transmittance spectra of one of the hole array designed
for transmission at 1550\,nm. This shows the dependencies of the
peaks width with the hole diameters. Peaks' heights are normalized
by factors shown in the legend. The small fast oscillations
correspond to Fabry-Perrot effects at the glass substrate
interfaces.} \label{spectrum}
\end{figure}

\begin{figure*}[htbp]
\begin{center}
\includegraphics[width=0.33\textwidth]{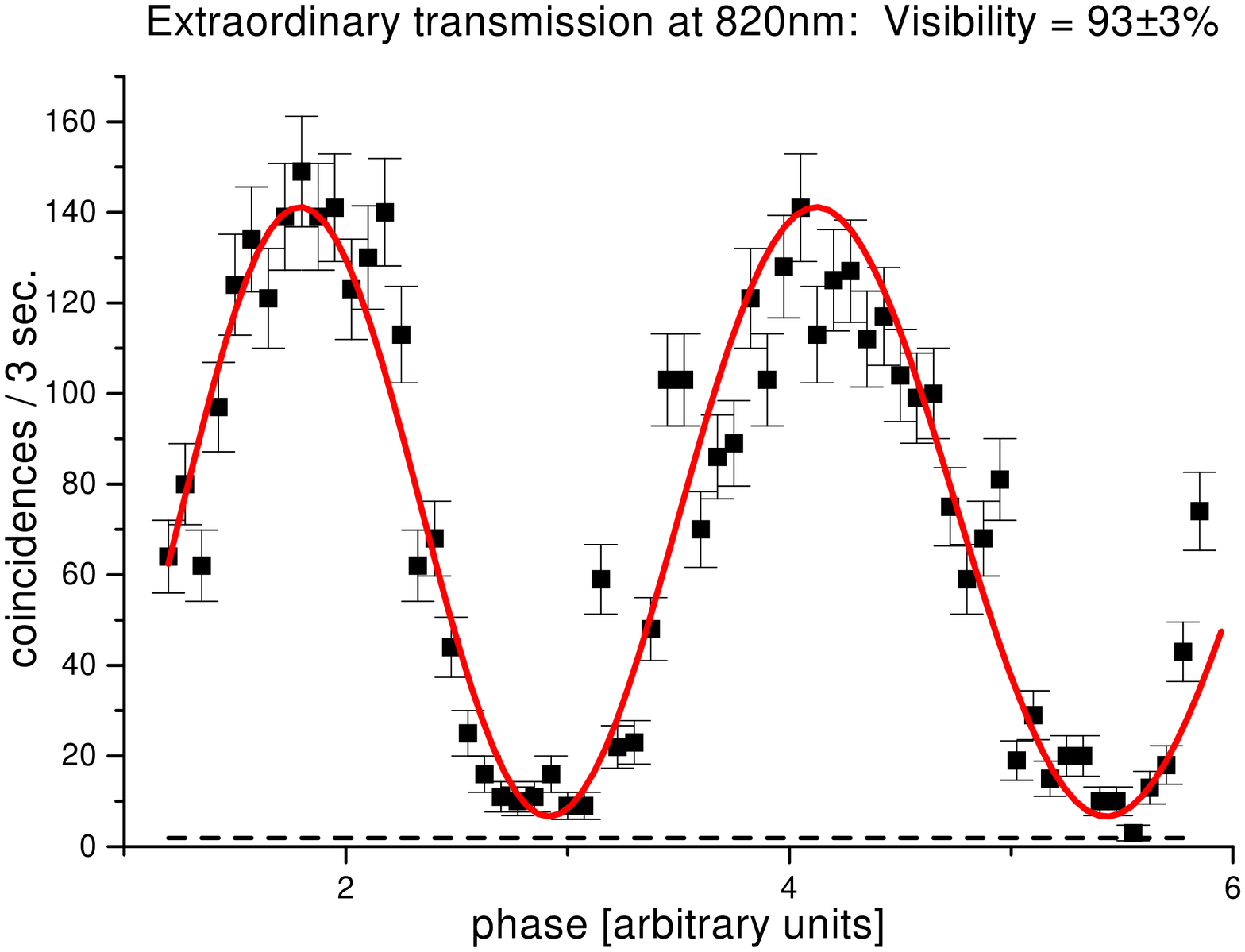}\includegraphics[width=0.33\textwidth]{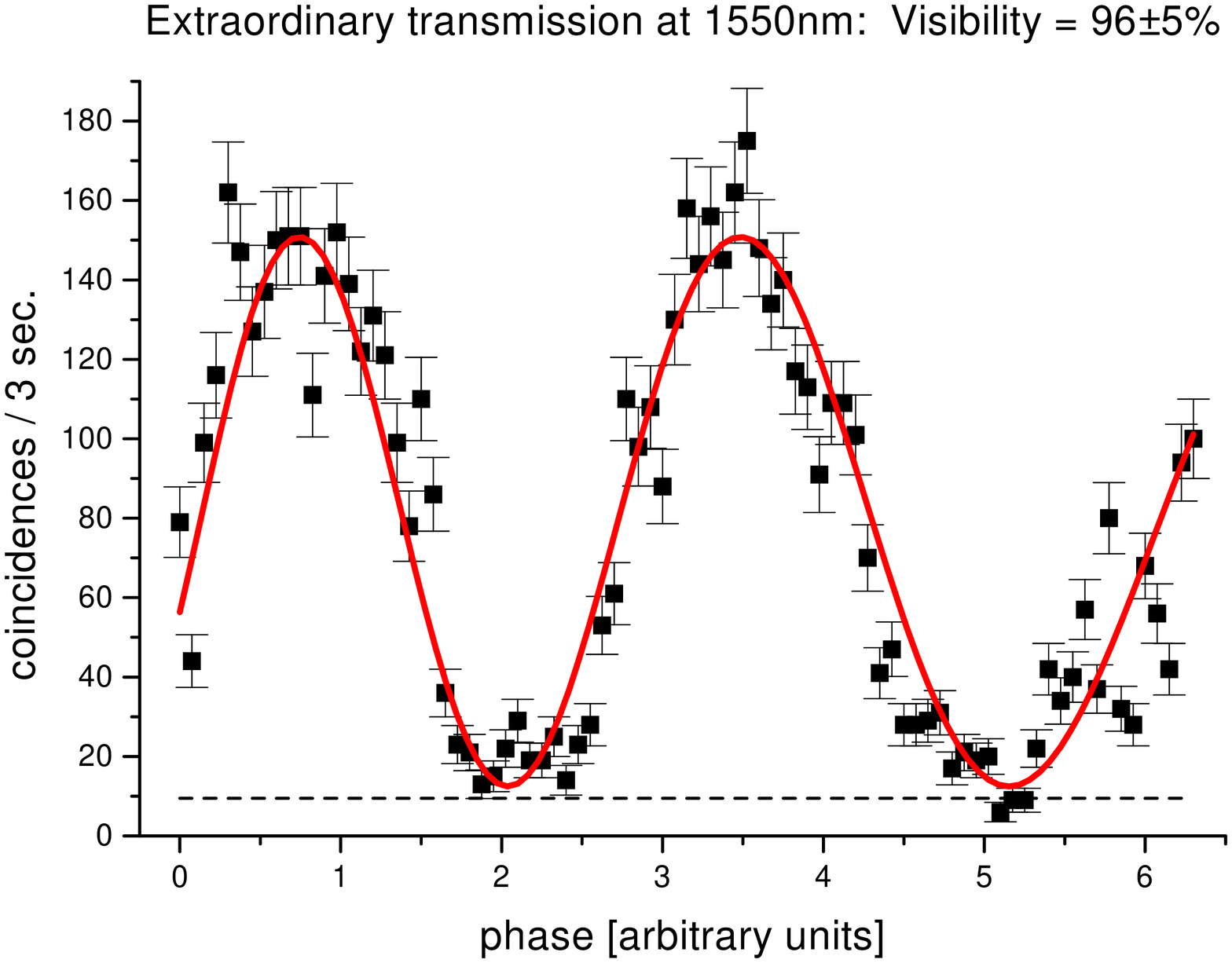}\includegraphics[width=0.33\textwidth]{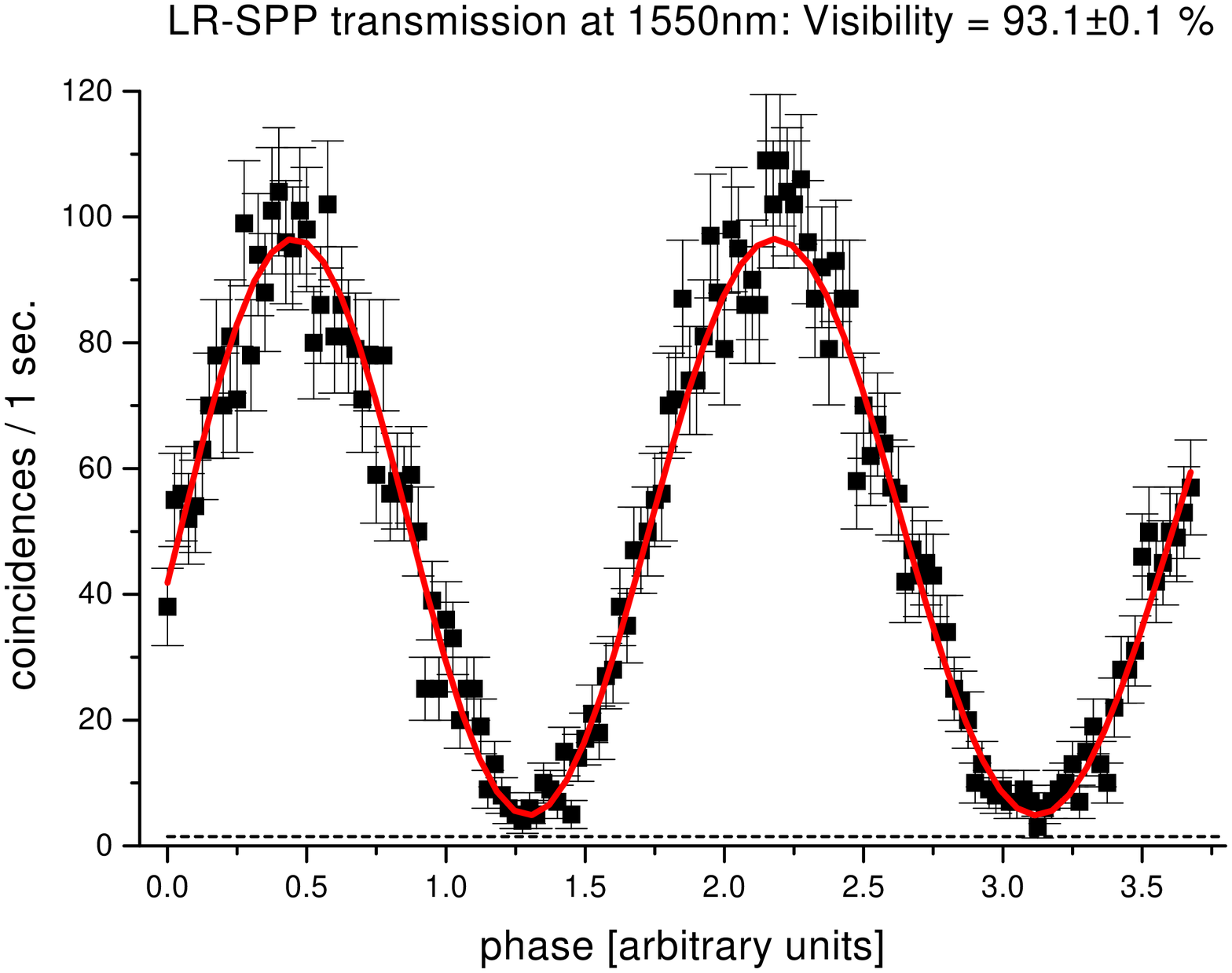}
\caption{Interference fringes with gold film samples and LR-SPP
waveguide in the path of the photons. The dashed horizontal line
is the noise level. The displayed visibilities are net value.}
\label{fringes}
\end{center}
\end{figure*}

In our samples, the extraordinary optical transmission is mediated
by SP mode lying at the metal-substrate interface and propagating
along the square array diagonals. In order to determine the
optimal periodicity $a$ and hole diameter $d$ for maximal
transmittance enhancement (i.e. for optimal coupling of the SPs to
the photon pair's members at the corresponding wavelengths)
various simulations were performed employing a modal expansion
method in connection with surface impedance boundary conditions on
the metal interfaces, and perfect metal boundary conditions in the
hole walls \cite{theoentang}. Several hole arrays were fabricated,
with slight changes around theoretical values and for a range of
fabrication parameters. They were characterized for transmittance
spectra, using classical light spectrometry. Results exhibit the
typical shape of SPs extraordinary light transmission, featuring
resonances due to SPs excitations (see figure \ref{spectrum}).
Only the spectra of the samples designed for the 1550\,nm
experiment are shown. To the best of our knowledge this
transmittance spectrum for extraordinary light transmission around
1550\,nm is the first reported so far. Spectra for the samples
designed for the 810\,nm have similar shapes. Verifications were
made to confirm that the re-emitted light originates from
photon-plasmon-photon conversions. First, the resonance position
scales linearly with the array period whereas the peaks height and
width depend on the hole diameter. Second, the SP band structure
was recovered by changing the polarization and incident angle of
the impinging photons. Results were very similar to the one
usually found in literature on the subject
\cite{extra,moreextra,moreextra2,polentang}.
From these measurements, the most suited arrays with respect to
our experimental limitations were chosen for the entanglement
experiment, one for the 810\,nm and one for the 1550\,nm photons.
The exact operating wavelengths $\lambda$ were chosen accordingly.
Entanglement measurements for the two wavelengths were made
separately. First, the source was tuned to produce photons at
820\,nm and 1515\,nm and a hole array with a=700\,nm and d=300\,nm
was used. Secondly, the source was tuned to produce photons at
810\,nm and 1550\,nm and a hole array with a=1400\,nm and
d=600\,nm was used. Interferences fringes are recorded without
samples inside the U-benches. The chosen sample is then inserted
inside the U-bench, and its position is adjusted using red laser
light. Interference fringes are then recorded again, and the ratio
of the maximal count rate in both cases is verified to be
compatible with the transmittance values previously measured with
classical light (see table \ref{results}). The resulting
interferences fringes are shown on figure \ref{fringes}.

\begin{figure}[htbp]
\includegraphics[width=\columnwidth]{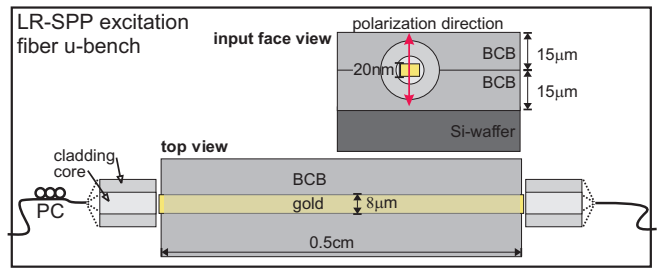}
\caption{Schematic of the fiber U-bench used to excite LR-SPP
propagation through gold stripe waveguide. }. \label{setupSPP}
\end{figure}

Using the same Franson-type experimental setup, we performed
another kind of measurement. The gold hole array at 1550\,nm is
replaced by a long range surface plasmon polariton (LR-SPP)
waveguide, provided by Micro Managed Photons A/S. This waveguide
consists of a 0.5\,cm long gold stripe), sandwiched between two
layers of benzocyclobutene (BCB), a dielectric of refraction index
$n=1.535$, and deposited upon a silicon waffer. Note that the
plasmon waveguide's length is much longer than the single photon
coherence length, hence during a certain time the photons are
entirely converted into plasmons. The gold stripe is 8\,$\mu$m
wide and about 20\,nm thick. Standard single mode fibers are
approached as close as possible from the ends of a thin gold strip
layer. The polarization mode is set to be linear and perpendicular
to the surface sample by mean of a fiber polarization controller.
SPP propagate along the gold stripe and re-emit light which is
collected by another single mode fiber (see figure
\ref{setupSPP}). This U-bench is connected in place of the
1550\,nm bulk U-bench of figure \ref{setup}, and interference
fringes are recorded (see figure \ref{fringes}). More details
about this gold stripe device and this particular plasmonic
excitation procedure can be found in reference \cite{lrspp} and
references therein.

The net visibilities of the interferences fringes are calculated
by substracting the noise level and fitting the data to a
sinusoidal dependency. The noise level is the contribution of the
InGaAs APD dark counts together with the double pair emission rate
of the SPDC source. The table \ref{results} show the obtained
values (visbilities are dark-count substracted net values).
\begin{table*}[htbp]
\caption{Experimental results}
\label{results}
\begin{ruledtabular}
\begin{tabular}{lccc}
Experiment&Reference visibility&Plasmon-assisted visibility&Transmittance\\
\hline
extraordinary transmission at 810\,nm   &$93\pm3$\%          &$93\pm3$\%                 &11\%\\
extraordinary transmission at 1550\,nm  &$97\pm3$\%          &$96\pm5$\%                 &6\%\\
LR-SPP transmission at 1550\,nm         &$93.1\pm0.1$\%      &$93.2\pm0.1$\%             &20\%
\end{tabular}
\end{ruledtabular}
\end{table*}
From these results, one can see that the strength of the quantum
correlations and thus the entanglement is clearly preserved by the
photon--plasmon--photon conversion, at the two working
wavelengths. No relevant two-photons interference visibility
reductions occur when a sample is inserted in the path of a photon
beam. The only effect is the reduction of the coincidence count
rate due to the partial transmittances resulting from
extraordinary light transmission. Energy-time entanglement is thus
proven to survives plasmon conversions at telecom wavelengths. The
preservation of this kind of entanglement implies that SPs are
coherently created by photons being in a superposition of two
different incoming times, separated by several nanoseconds, while
the SPs lifetime in our experiment should be much less than
picoseconds (the time for light to travel through the sample
thickness). Therefore the only SP quantum state compatible with
the above results is a superposition of a single SP existing at
two different moment in time separated one from the other by a
duration thousand times longer than the its own lifetime. At a
macroscopic level this would lead to a "Schrödinger cat" living at
two epochs that differ by much more than a cat's lifetime. The
plasmons of the present experiment involve a mesoscopique number
of electrons of about $10^{10}$. It should however be stressed
that these do carry collectively only a single degree of freedom,
i.e. a single qubit.

Apart from these fundamental considerations, our results show that
energy-time entanglement can be efficiently coupled to SPs. As
this kind of entanglement is robust for long distance quantum
information transmission, future quantum SPs circuits could be
integrated with long range optical quantum communication networks.

\begin{acknowledgments}
Micro Managed Photons (MMP) A/S (Denmark) is greatfully
acknowledged for providing the gold stripe LR-SPP samples together
with some data and expert advices. Financial support by the Swiss
NCCR Quantum Photonics is acknowledged.
\end{acknowledgments}


\begin{thebibliography}{30}
\bibitem{robust} R. T. Thew, S. Tanzilli,. Tittel, H. Zbinden and N. Gisin, Phys. Rev. A {\bf 66}, 062304 (2002)
\bibitem{cryptoreview} N. Gisin, G. Ribordy, W. Tittel, and H. Zbinden, Rev. Mod. Phys. {\bf 74}, 145 (2002)
\bibitem{teleportbase} C. H. Bennett, G. Brassard, C. Crépeau, R. Jozsa, A. Peres, and W. K. Wootters, Phys. Rev. Lett. {\bf 70}, 1895 (1993)
\bibitem{monroe} C. Monroe, Nature {\bf 416}, 238 (2002)
\bibitem{polzik2} K. Hammerer, K. Molmer, E.S. Polzik, J.I. Cirac, quant-ph/0312156
\bibitem{barrett} M. D. Barrett et al., Nature {\bf 429}, 737 (2004)
\bibitem{polzik3} J. Sherson, B. Julsgaard, E. S. Polzik, quant-ph/0408146
\bibitem{plasmonsreview} W. L. Barnes, A. Dereux, and T. W. Ebbesen, Nature {\bf 424}, 824 (2003)
\bibitem{bethe} H. A. Bethe, Phys. Rev. {\bf 66}, 163 (1944)
\bibitem{extra} T. W. Ebbesen et al., Nature {\bf 391}, 667 (1998)
\bibitem{moreextra} H. F. Ghaemi, T. Thio, D. E. Grupp, T. W. Ebbesen, and H. J. Lezec, Phys. Rev. B {\bf 58}, 6779 (1998)
\bibitem{moreextra2} D. E. Grupp, H. J. Lezec, T. W. Ebbesen, K. M. Pellerin, and T. Thio, Appl. Phys. Lett. {\bf 77}, 1569 (2000)
\bibitem{theoextra} L. Martín-Moreno et al., Phys. Rev. Let. {\bf 86}, 1114 (2001)
\bibitem{polentang} E. Altewischer et al., Nature {\bf 418}, 304 (2002)
\bibitem{theoentang} E. Moreno, F. J. García-Vidal, D. Erni, J. I. Cirac, and L. Martín-Moreno, Phys. Rev. Lett. {\bf 92}, 236801 (2004)
\bibitem{franson} J. D. Franson, Phys. Rev. Lett. {\bf 62}, 2205 (1989)
\bibitem{cryptoribordy} G. Ribordy, J. Brendel, J.-D. Gautier, N. Gisin, and H. Zbinden, Phys. Rev. A {\bf 63}, 012309 (2000)
\bibitem{cryptofasel} S. Fasel, G. Ribordy, H. Zbinden and N. Gisin, Eur. Phys. J. D {\bf30}, 143 (2004)
\bibitem{lrspp} T. Nikolajsen, K. Leosson, I. Salakhutdinov, and S. I Bozhevolnyi, Appl. Phys. Lett. {\bf 82}, 668 (2003)
\end{thebibliography}
\end{document}